**How Flexoelectricity Drives Triboelectricity.**

C. A. Mizzi and L. D. Marks[1]
Department of Materials Science and Engineering
Northwestern University, Evanston, IL 60208

**Corresponding Author**

[1]To whom correspondence should be addressed. Email: L-marks@northwestern.edu.

**Abstract**

Triboelectricity has been known since antiquity, but the fundamental physics underlying this phenomenon lacks consensus. We present a flexoelectric model for triboelectricity where contact deformation induced band bending is the driving force for charge transfer. This framework is combined with first principles and finite element calculations to explore charge transfer implications for different contact geometry and materials combinations. We demonstrate that our *ab initio* based formulation is compatible with existing empirical models and experimental observations including charge transfer between similar materials and size/pressure dependencies associated with triboelectricity.

Triboelectricity, the charge transfer associated with contacting and rubbing two materials, has been known for at least twenty-five centuries [1,2]. It can be a boon or a bane in industries [3] ranging from xerography [4] to pharmaceuticals [5], and is thought to play a critical role in many processes such as dust storms [6] and planetary formation [7]. To date a consensus on the fundamental physics of triboelectricity including the driving force and transferred charged species has proven elusive. For metal-on-metal contact it is generally accepted that charging is in part associated with electron transfer driven by contact potential differences [8]. The literature is less-clear for metal-on-insulator and insulator-on-insulator contacts, with experiments suggesting a variety of charged species [1-3,9] and models invoking several mechanisms such as temperature differences [10] and trapped charge [11,12].

Long before Bowden and Tabor established asperity contact as the origin of friction and wear [13], Volta and Helmholtz [14-16] hypothesized that the role of rubbing in tribocharging was to increase the number of contact points between two materials. Since then it has been shown that *macroscopic* deformations play an important role in triboelectricity, first in 1910 when bending was found to dictate the direction of charge transfer [17], with further work over the next century [18-21], but little attention has been given to the *microscopic* deformations at asperity contacts. We recently analyzed these deformations using flexoelectricity [22], the electromechanical coupling between polarization and strain gradient present in all insulators [23]. We pointed out that for typical rubbing or contact conditions, elastic strain fields at contacting asperities would generate potential differences via the flexoelectric effect that are large enough to drive charge transfer. This analysis, which contained no adjustable parameters beyond material properties, was demonstrated to be consistent with many experimental observations including charge mosaics [24] and bipolar charging during stick-slip [25], and an independent expansion of this model to randomly rough surfaces has also shown good agreement with experiment [26].

In this Letter, we extend and generalize the flexoelectric model for triboelectric charging. We construct interfacial band diagrams for different combinations of contact geometries and materials accounting for deformation-induced changes to electronic structure. This analysis suggests that flexoelectricity-driven band bending during contact between different/like materials is a mechanical analogue of traditional band bending in hetero/homo-junctions in semiconductor devices; transfer of charge during contact can be thought of as a contact-driven diode. Our findings are consistent with existing empirical models involving trap states, contact potentials, and ion transfer, but now these dependencies arise naturally from an *ab initio* based formulation of the problem. Of particular importance, the results indicate that depending upon both material and force the system can act as either a conventional or a Zener/breakdown diode, which explains variations in both the sign and magnitude of charge transfer with force which has led to confusing and contradictory experimental results, and impeded progress for decades.

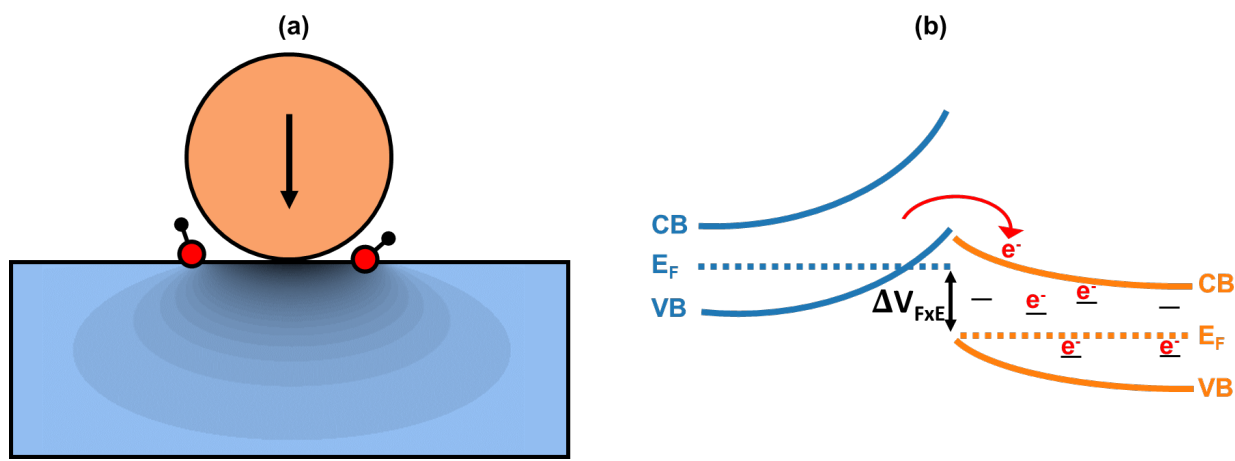

Figure 1. (a) Schematic of sphere-on-flat contact depicting elastic deformation fields emanating from the contact point and adsorbed species on the surface. (b) Elastic deformations induce an additional contact potential term through flexoelectric couplings. Depending on the materials, geometry, and extent of deformation, contact-induced band bending can drive the occupation of trap states, bulk charge transfer, or the adsorption of charged species.

As illustrated in Figure 1, elastic deformations induce band bending around a contact point. A description of such electronic structure changes (within the Schottky-Mott limit of no charge transfer [27,28]) corresponds to determining the spatial variation of the conduction band (CB) and valence band (VB) edges in each contacting body with deformation. In a centrosymmetric insulator subjected to an inhomogeneous strain, the energy of a band feature $E_i$ referenced to vacuum $E_{vac}$ will vary according to

$$\left(E_i(\mathbf{r}) - E_{vac}(\mathbf{r})\right)_{strained} - \left(E_i(\mathbf{r}) - E_{vac}(\mathbf{r})\right)_{unstrained} = \Delta \bar{V}_{FxE}(\mathbf{r}) + (\varphi + D_i)\epsilon(\mathbf{r}) \quad (1)$$

The first term $\Delta \bar{V}_{FxE}(\mathbf{r})$ corresponds to the change in the average Coulomb potential (also known as the mean-inner potential) arising from strain gradients via the bulk flexoelectric effect. It is given by

$$\Delta \bar{V}_{FxE}(\mathbf{r}) = \bar{V}(\mathbf{r}) - \bar{V}_0 = \frac{1}{4\pi\varepsilon_0(1+\chi)} \int_\Omega \frac{\mathbf{P}_{FxE}(\mathbf{r}') \cdot (\mathbf{r} - \mathbf{r}')}{|\mathbf{r} - \mathbf{r}'|^3} d\Omega' \quad (2)$$

where $\bar{V}(\mathbf{r})$ is the mean-inner potential at point $\mathbf{r}$ in the deformed body referenced to $\bar{V}_0$, the mean-inner potential sufficiently far from the contact point, $\varepsilon_0$ is the permittivity of free space, $\chi$ is the dielectric susceptibility, and $\mathbf{P}_{FxE}(\mathbf{r}')$ is the flexoelectric polarization at point $\mathbf{r}'$. The second term $\varphi$ gives the shift in mean-inner potential with respect to vacuum (also known as the surface potential offset) induced by strain $\epsilon$.

$$\varphi = \frac{d(\bar{V} - E_{vac})}{d\epsilon} \quad (3)$$

This term enables a direct comparison between different materials by introducing an absolute energy reference and implicitly includes "surface" flexoelectric effects [29]. The third term $D_i(\mathbf{r})$ describes the relative deformation potential, i.e. the local change in energy of a band feature relative to the mean-inner potential from strain [30,31].

$$D_i = \frac{d(E_i - \bar{V})}{d\epsilon} \qquad (4)$$

The relative offset between band features in two materials in contact can be found by combining the above expressions with the work function ($\phi$) difference between the two materials.

To explore the implications of Eq. (1) for specific cases, we make three simplifications. First, we consider contact geometries with axial symmetry between isotropic materials so the non-trivial flexoelectric polarization components entering Eq. (2) within a cylindrical coordinate system are

$$P_r(r,z) = \mu_L \epsilon_{rr,r}(r,z) + \mu_T\big(\epsilon_{zz,r}(r,z) + \epsilon_{\theta\theta,r}(r,z)\big) + 2\mu_S \epsilon_{rz,z}(r,z) \qquad (5)$$

$$P_z(r,z) = \mu_L \epsilon_{zz,z}(r,z) + \mu_T\big(\epsilon_{rr,z}(r,z) + \epsilon_{\theta\theta,z}(r,z)\big) + 2\mu_S \epsilon_{rz,r}(r,z) \qquad (6)$$

$\epsilon_{ij,k} = \frac{\partial \epsilon_{ij}}{\partial x_k}$ where $\epsilon_{ij}$ is symmetrized strain, $P_r$ and $P_z$ are radial and axial components of the polarization, and $\mu_L$, $\mu_T$, and $\mu_S$ are longitudinal, transverse, and shear bulk flexoelectric coefficients. Second, we only treat volumetric strain effects in Eq. (3) and (4) as shear tends to split, not shift, energy levels [32]. Third, elastic contact is assumed, which is likely a lower bound on flexoelectric contributions to triboelectricity: plastic deformation would have additional contributions from point [33], line [34], and planar [35] defects which are known to enhance a flexoelectric response. Materials parameters for Eq. (1), i.e. flexoelectric coefficients, volumetric deformation potentials, and surface potential offsets, are calculated with density functional theory (Supplemental Material [36]), but these values could readily come from experiments. Deformation fields are determined from finite element calculations (Supplemental Material [36]) to avoid errors associated with analytic approximations (e.g. Hertz theory [37] incorrectly predicts identical deformations for contact between chemically identical bodies with different curvatures, Supplemental Material [36]).

Figure 2 compares the relative strengths of the three terms in Eq. (1) for the CB minimum in a flat SrTiO$_3$ sample deformed by a sphere with a contact pressure of 8 GPa. This contact pressure corresponds to the hardness of SrTiO$_3$ [38], which is an order-of-magnitude elastic limit for the maximum contact pressure [39]. The deformation potential and surface potential offset terms are largest near the contact point, decay rapidly with distance, and mostly cancel (although this is material dependent). In contrast the flexoelectric term is concentrated near the contact radius, but remains significant throughout the contact volume. The net effect is inhomogeneous band bending of ~ $\pm 1$ V within the vicinity of the contact area, even for soft contact involving materials with modest flexoelectric properties such as SrTiO$_3$. The magnitude of band bending within the contact radius will monotonically increase with contact pressure (Supplemental Material [36]).

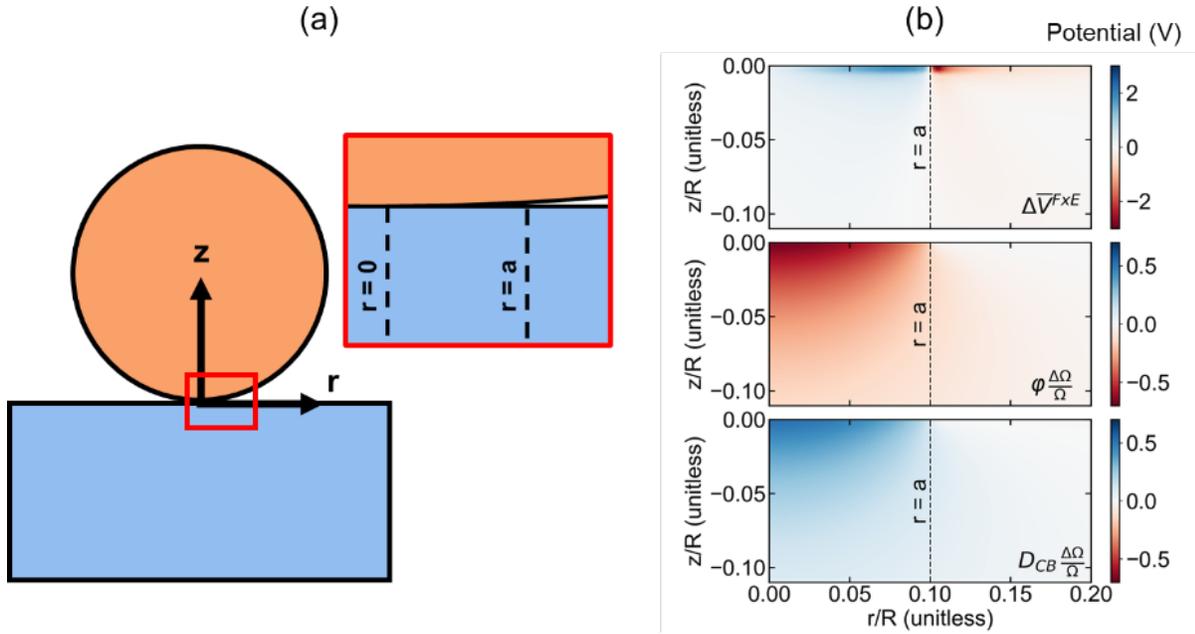

Figure 2. (a) Schematic of sphere-on-flat contact with coordinate system definition. (b) Change in the mean-inner potential owing to flexoelectricity ($\Delta \bar{V}^{FxE}$), the mean-inner potential relative to vacuum ($\varphi \frac{\Delta\Omega}{\Omega}$), and the conduction band minimum relative to the mean-inner potential ($D_{CB}\frac{\Delta\Omega}{\Omega}$) in flat SrTiO$_3$ contacted by a sphere with a contact pressure of 8 GPa. Distances are normalized by the indenter radius R.

We now utilize the band bending framework developed above to construct interfacial band diagrams for sphere-on-flat contact and analyze the implications for charge transfer. First, we focus on contact between dissimilar materials using a Si sphere and SrTiO$_3$ flat as an example. Plots of band bending at constant pressure and different radial distances from the contact point are shown in Fig. 3, and at different contact pressures at a fixed radial distance in Fig. 4. Spatially inhomogeneous band bending is observed in both bodies with the largest band bending occurring in the vicinity of the contact radius. The spatial evolution of the band bending depends sensitively on the flexoelectric coefficients; for example, there are qualitative differences in the band bending profiles of Si and SrTiO$_3$ because Si has larger shear contributions than SrTiO$_3$ (Supplemental Material [36]).

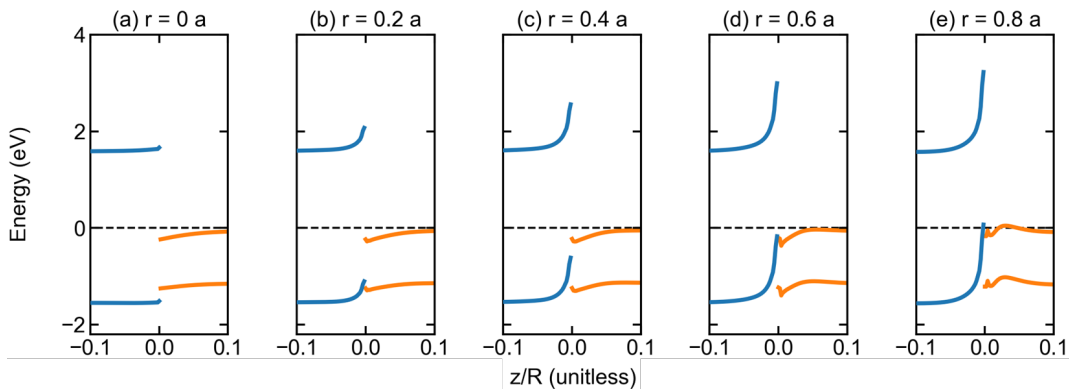

Figure 3. Sphere-on-flat contact band diagrams for a Si sphere (orange) on a SrTiO$_3$ flat (blue) including flexoelectric, deformation potential, and surface potential offset effects, and a work function difference of 0.6 eV. (a)-(e) show the conduction and valence band edges as a function of depth (normalized by the indenter radius R) at different radial distances (in units of the contact radius a) from the contact point as defined in Fig. 2(a) with a contact pressure of 6 GPa. The unstrained Fermi level of each material is assumed to be at its band gap center and zero energy is taken to be the unstrained SrTiO$_3$ Fermi level.

Traditional contact potential theory used to describe metal-on-metal tribocharging [8] is often deemed inapplicable to insulators because band gaps present too large an energy barrier for charge transfer [9,14,40]. Our model generalizes the theory of contact potential differences, reproducing the well-documented result that work function differences alone are insufficient to explain charge transfer in insulators, while demonstrating that band bending from elastic deformations lowers the energy barrier for transferring charge from one insulating body to another. In the limit of no in-gap states our model predicts zero charge transfer below a contract pressure threshold (e.g. Fig. 4(a)) followed by some transfer assuming states are available (e.g. hole transfer from the SrTiO$_3$ VB to the Si VB of in Fig. 4(b)-(c)). Sufficiently hard contact yields direct transfer between bulk states on different bodies via Zener-tunneling [41]. In the example shown in Fig. 4(e) this corresponds to electrons tunneling from the VB of SrTiO$_3$ into the CB of Si around the edge of the contact.

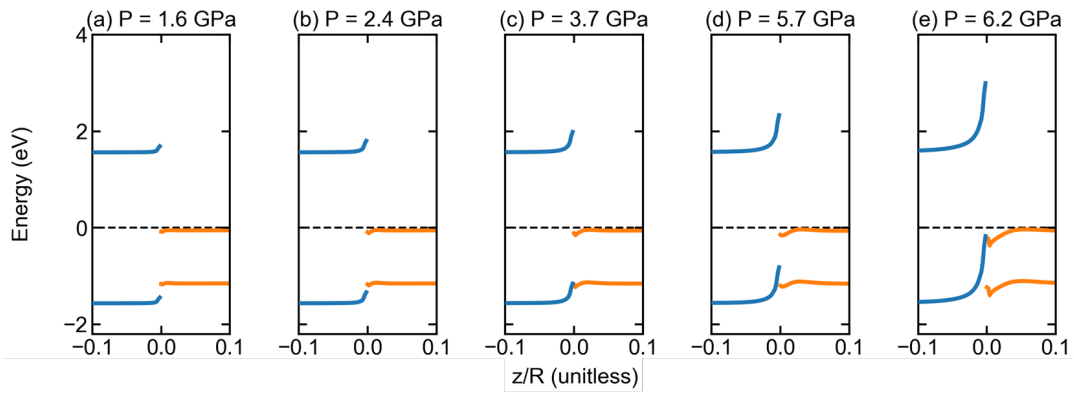

Figure 4. Sphere-on-flat contact band diagrams for a Si sphere (orange) on a SrTiO$_3$ flat (blue) including flexoelectric, deformation potential, and surface potential offset effects, and a work function difference of 0.6 eV. (a)-(e) show the conduction and valence band edges as a function of depth (normalized by the indenter radius R) at different contact pressures for a fixed radial distance of 0.6a from the contact point. The unstrained Fermi level of each material is assumed to be at its band gap center and zero energy is taken to be the unstrained SrTiO$_3$ Fermi level.

In a fully adiabatic limit the aforementioned charge transfer would be reversible, however this limit is not often realized. In practice there will always be in-gap states either at defects or surfaces that trap charge so that when the two bodies return to their original configurations after contact, charge within trap states remains leaving the materials charged. The existence of trapped charged and its relationship to triboelectricity is directly supported by tribo-luminescence experiments [42,43], and the longevity of charge in trap states is well-documented [44]. Trap states have long been suggested to play an important role in triboelectricity and some triboelectric models involve the redistribution of non-equilibrium charge distributions in localized trap states during contact [11,12]. However, the origin of the non-equilibrium distributions in these models has been an open question and experiments suggest there is insufficient intrinsic trapped charge to explain tribocharging [45]. Our work indicates non-equilibrium charge distributions are a natural consequence of contact and that contact deformations may significantly increase the trapped charge densities beyond intrinsic amounts. The underlying physics in this model is analogous to increases in trapped charge densities from applied potentials observed at semiconductor device interfaces [46], except here trap occupation is mechanically driven.

Now we turn to sphere-on-flat contact between two $SrTiO_3$ bodies. The identical material case shown in Fig. 5 is important because it models the minimal asymmetry between two bodies in which one should expect tribocharging [10], and is directly relevant for many cases such as in dust storms [6]. We note that in the Hertzian limit, band bending during contact between two chemically identical bodies is symmetric for all combinations of curvature so there will be no contact-deformation-driven charge transfer for two defect-free bodies (Supplemental Material [36]). Beyond Hertz theory there are subtle but important differences between two bodies with different curvatures, although the differences are smaller than for dissimilar materials. Fig. 5 shows asymmetric band bending across the contact interface with the largest differences around the edge of the contact region, similar to the dissimilar material case in Fig. 3. For the conditions used in this simulation there would be some transfer of electrons from the sphere to the flat, assuming that there were states available. Like the case involving dissimilar materials, in a fully adiabatic limit this charge transfer would be reversible but with traps present it will not be. This result provides an explanation for the experimental observation that larger bodies tend to charge positive with respect to smaller bodies made of the same material [45], however our model predicts the direction of charge transfer is dictated by the relative size of the bodies and their flexoelectric coefficients.

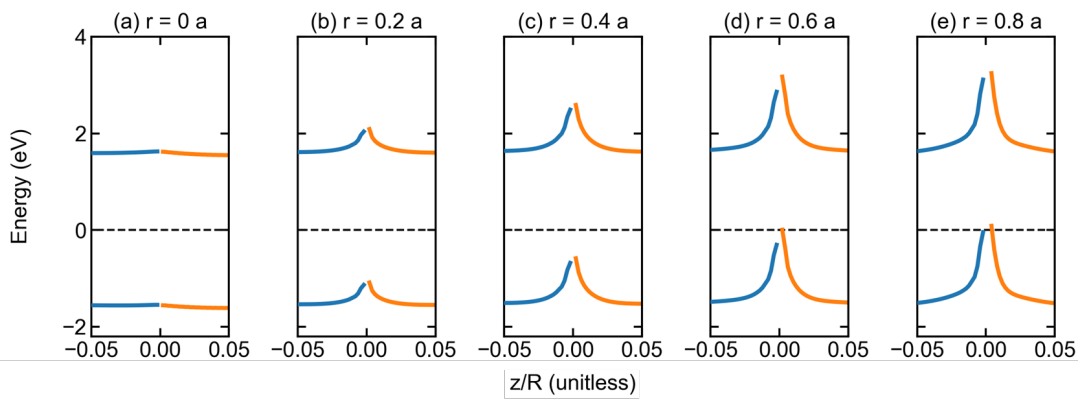

Figure 5. Sphere-on-flat contact band diagrams for a SrTiO$_3$ sphere (orange) on a SrTiO$_3$ flat (blue) including flexoelectric, deformation potential, and surface potential offset effects. (a)-(e) show the conduction and valence band edges as a function of depth (normalized by the indenter radius R) at different radial distances (in units of the contact radius a) from the contact point as defined in Fig. 2(a) with a contact pressure of 8 GPa. The unstrained Fermi level is assumed to be at the band gap center and taken to be zero energy.

In the above we have focused on charge transfer mechanisms involving electrons, however contact-induced band bending also has implications for ion transfer. There is experimental evidence that ions [1,2,9], often OH$^-$ or H$^+$ [47-49], are exchanged during the tribocharging of insulators. The interplay between band bending and adsorbed species is also well-established [50] and it is known that band bending at surfaces [50] can drive adsorption [51]. Likewise, band bending originating from contact deformations in our model can serve as a driving force for ion transfer, although the relevant band banding for this mechanism would be outside the contact radius (Supplemental Materials [36]). This process would be largely irreversible because ionic motion is slow on the time scale of contact [14].

Definitive matching of the band bending we predict and resultant charge transfer in experiments requires finer control of the number, type, and energy of defect states than is typical of triboelectric experiments. However, we argue that our application of established physics (flexoelectricity, band bending, elasticity) explains many existing experimental results at a qualitative or semi-quantitative level; besides the connections to contact potential theory, trap states, and electron/ion transfer already described in this Letter, we provide a few additional examples here and more in the Supplemental Materials [36]. First, many experiments have shown that the sign of transferred charge changes with pressure, e.g. [16,52]. This is captured within our model by the pressure-dependent onset of Zener tunneling. Next, temperature dependencies are implicitly included in this model through the statistical occupation of states. Given its formulation as a band theory, this model can also be readily expanded to include other temperature effects related to band structure such as Thomson/Seebeck effects that may be relevant during rubbing [10,40,53]. Additionally, this model explains experiments involving band bending associated with conductive atomic force microscopy experiments [54,55]. This technique, possibly in conjunction with optical methods, is likely the optimal method to test our model as it provides a direct and local measure of electromechanical couplings.

The model and simulations presented in this Letter demonstrate the existence of sizable inhomogeneous band bending during contact owing to flexoelectricity. Consequences of this model are consistent with experimental observations and provide a unified, *ab initio* driver for common charge transfer mechanisms. Though the detailed charge transfer mechanisms will be complex and sensitive to defects, adsorbates, etc., the inevitability of large deformation during contact and the universality of flexoelectricity make considerations of these contributions to triboelectricity important and necessary. The framework developed here is amenable to arbitrary materials/geometries and makes sizable progress towards *a priori* predictions of triboelectric charge transfer between insulators.

**Acknowledgements**

This work was supported by the National Science Foundation (NSF) under grant number DMR-1507101 and the U.S. Department of Energy, Office of Science, Basic Energy Sciences, under Award No. DE-FG02-01ER4594. C.A.M. performed the analysis supervised by L.D.M. Both authors contributed to the writing of the Letter.

**Supplemental Material: How Flexoelectricity Drives Triboelectricity.**
C. A. Mizzi and L. D. Marks[1]
Department of Materials Science and Engineering
Northwestern University, Evanston, IL 60208

**Corresponding Author**

[1]To whom correspondence should be addressed. Email: L-marks@northwestern.edu.

**S1. Flexoelectric contact model.**

*Flexoelectric potential.*

Here we derive an expression for the change in the mean inner potential (MIP) arising from a flexoelectric polarization. In the absence of back-couplings, the phenomenological expression for polarization $P_i$ in a centroysmmetric insulator with dielectric and flexoelectric couplings [1,2] is

$$P_i = \varepsilon_0 \chi_{ij} E_j + P_i^{FxE} \quad (1)$$

where $\varepsilon_0$ is the permittivity of free space, $\chi_{ij}$ is the dielectric susceptibility, $E_j$ is the electric field, and $P_i^{FxE}$ is the flexoelectric polarization. Each component of $P_i^{FxE}$ is given by

$$P_i^{FxE} = \mu_{ijkl} \frac{\partial \epsilon_{kl}}{\partial x_j} \quad (2)$$

where $\mu_{ijkl}$ is the (zero electric field) flexoelectric coefficient tensor and $\frac{\partial \epsilon_{kl}}{\partial x_j}$ is the strain gradient. We use the symmetrized strain convention such that $\epsilon_{ij} = \frac{1}{2}\left(\frac{\partial u_i}{\partial x_j} + \frac{\partial u_j}{\partial x_i}\right)$, $\epsilon_{ij} = \epsilon_{ji}$, and $\frac{\partial \epsilon_{ij}}{\partial x_k} = \frac{\partial \epsilon_{ji}}{\partial x_k}$, and note that in Eq. (1) and (2), there is an implicit summation over repeated indices.

The electric field is related to the total charge density $\rho$ by

$$\vec{\nabla} \cdot \vec{E} = \frac{\rho}{\varepsilon_0} = \frac{\rho_b + \rho_f}{\varepsilon_0} \quad (3)$$

where we have written the total charge density as the sum of the bound charge ($\rho_b$) and free charge ($\rho_f$) on the right. The bound and free charges are related to divergences of the polarization and dielectric displacement, respectively, by the following two equations:

$$\vec{\nabla} \cdot \vec{P} = -\rho_b \tag{4}$$

$$\vec{\nabla} \cdot \vec{D} = \rho_f \tag{5}$$

Assuming a perfect insulator such that $\rho_f = 0$ and a diagonal dielectric susceptibility tensor with a single eigenvalue $\chi$, Eq. (1) – (5) can be used to relate the divergence of the electric field to the divergence of the flexoelectric polarization yielding

$$\vec{\nabla} \cdot \vec{E} = -\frac{1}{\varepsilon_0(1+\chi)}(\vec{\nabla} \cdot \vec{P}_{FxE}) \tag{6}$$

Eq. (6) can be also be written in terms of the electrostatic potential as

$$\nabla^2 V = \frac{1}{\varepsilon_0(1+\chi)}(\vec{\nabla} \cdot \vec{P}_{FxE}) \tag{7}$$

which is a Poisson equation with a charge density associated with the flexoelectric effect. Defining this charge density as

$$\rho_{FxE} = -\frac{1}{(1+\chi)}\vec{\nabla} \cdot \vec{P}_{FxE} \tag{8}$$

allows us to recover the Poisson equation in its typical form

$$\nabla^2 V = -\frac{1}{\varepsilon_0}\rho_{FxE} \tag{9}$$

This can be solved by identifying appropriate electrostatic potential boundary conditions or recasting the problem in integral form and choosing an appropriate reference. In the latter case, an equivalent integral form to the differential form given above [3] is

$$V(\vec{r}) = \frac{1}{4\pi\varepsilon_0(1+\chi)} \left( \int_\Omega \frac{\vec{P}_{FxE}(\vec{r}') \cdot (\vec{r}-\vec{r}')}{|\vec{r}-\vec{r}'|^3} d\Omega' \right) \qquad (10)$$

Since only cases with axial symmetry are considered here, Eq. (10) reduces to

$$V(\vec{r}) = \frac{1}{2\,\varepsilon_0(1+\chi)} \left( \int \frac{\vec{P}_{FxE}(\vec{r}') \cdot (\vec{r}-\vec{r}')}{|\vec{r}-\vec{r}'|^3} \rho\, d\rho\, dz \right) \qquad (11)$$

where $\rho$ is the radial coordinate and z is the axial coordinate.

As described in detail in Ref. [4], bulk flexoelectric coefficients describe the long-range contributions to the absolute deformation potential [5] or equivalently the tilt of the mean inner potential under an inhomogeneous strain. This means that the flexoelectric potential in Eq. (10) and (11) implicitly references the value of the mean inner potential (MIP) in an undeformed crystal. Therefore we set this reference to be the MIP at the furthest point from the contact point. We also rewrite Eq. (11) to explicitly include the reference potential.

$$\Delta \bar{V}^{FxE} = \bar{V}(\vec{r}) - \bar{V}_0 = \frac{1}{2\,\varepsilon_0(1+\chi)} \left( \int \frac{\vec{P}_{FxE}(\vec{r}') \cdot (\vec{r}-\vec{r}')}{|\vec{r}-\vec{r}'|^3} \rho\, d\rho\, dz \right) \qquad (12)$$

To confirm our choice of reference potential is appropriate, Supplemental Fig. S1 shows the variation in the MIP owing to the bulk flexoelectric effect as a function of depth at the point of contact between a Si sphere and SrTiO$_3$ flat with a contact pressure of 6 GPa. The quick return to zero potential as a function of depth validates our choice of reference. Details on how this data was calculated is provided in the subsequent sections.

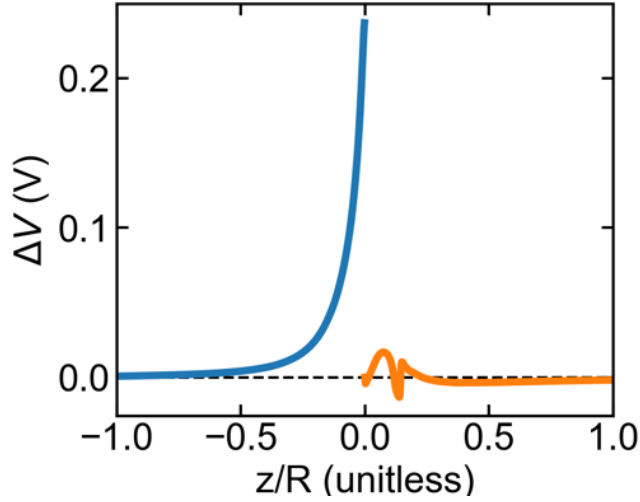

Supplemental Figure S1. Variation in the mean inner potential at the point of contact (r = 0) as a function of depth (z) normalized by the sphere radius (R) in a SrTiO$_3$ flat (blue) contacted by a Si sphere (orange). A contact pressure of 6 GPa is used. Qualitative differences in the two profiles are because of the relative size of the different flexoelectric coefficients in SrTiO$_3$ and Si (Section S6).

*Flexoelectric Polarization.*

To use Eq. (12) it is first necessary to obtain an expression for $\vec{P}_{FxE}$. In this work, cubic symmetry is assumed which reduces $P_x$ to

$$P_x = \mu_{xxxx}\frac{\partial \epsilon_{xx}}{\partial x} + \mu_{xyxy}\frac{\partial \epsilon_{xy}}{\partial y} + \mu_{xzxz}\frac{\partial \epsilon_{xz}}{\partial z} + \mu_{xyyx}\frac{\partial \epsilon_{yx}}{\partial y} + \mu_{xxyy}\frac{\partial \epsilon_{yy}}{\partial x} + \mu_{xzzx}\frac{\partial \epsilon_{zx}}{\partial z} + \mu_{xxzz}\frac{\partial \epsilon_{zz}}{\partial x} \quad (13)$$

To further reduce the expression above, we (1) note that according to the strain convention adopted in this work, $\frac{\partial \epsilon_{ij}}{\partial x_k} = \frac{\partial \epsilon_{ji}}{\partial x_k}$ and (2) introduce the notation $\mu_{xxxx} = \mu_L$, $\mu_{xxyy} = \mu_{xxzz} = \mu_T$, and $\mu_{xyxy} = \mu_{xzxz} = \mu_{xyyx} = \mu_{xzzx} = \mu_S$ which holds for a cubic material, where L, T, and S refer to longitudinal, transverse, and shear, respectively, so we can write

$$P_x = \mu_L \frac{\partial \epsilon_{xx}}{\partial x} + \mu_T \left(\frac{\partial \epsilon_{yy}}{\partial x} + \frac{\partial \epsilon_{zz}}{\partial x}\right) + 2\mu_S \left(\frac{\partial \epsilon_{xy}}{\partial y} + \frac{\partial \epsilon_{xz}}{\partial z}\right) \quad (14)$$

Similar expression can be derived for $P_y$ and $P_z$:

$$P_y = \mu_L \frac{\partial \epsilon_{yy}}{\partial y} + \mu_T \left( \frac{\partial \epsilon_{xx}}{\partial y} + \frac{\partial \epsilon_{zz}}{\partial y} \right) + 2\mu_S \left( \frac{\partial \epsilon_{yx}}{\partial x} + \frac{\partial \epsilon_{yz}}{\partial z} \right) \tag{15}$$

$$P_z = \mu_L \frac{\partial \epsilon_{zz}}{\partial z} + \mu_T \left( \frac{\partial \epsilon_{xx}}{\partial z} + \frac{\partial \epsilon_{yy}}{\partial z} \right) + 2\mu_S \left( \frac{\partial \epsilon_{zx}}{\partial x} + \frac{\partial \epsilon_{zy}}{\partial y} \right) \tag{16}$$

Combining this with the axial symmetry of the problem and adopting cylindrical coordinates yields

$$P_r = \mu_L \frac{\partial \epsilon_{rr}}{\partial r} + \mu_T \left( \frac{\partial \epsilon_{\theta\theta}}{\partial r} + \frac{\partial \epsilon_{zz}}{\partial r} \right) + 2\mu_S \left( \frac{\partial \epsilon_{rz}}{\partial z} \right) \tag{17}$$

$$P_z = \mu_L \frac{\partial \epsilon_{zz}}{\partial z} + \mu_T \left( \frac{\partial \epsilon_{\theta\theta}}{\partial z} + \frac{\partial \epsilon_{rr}}{\partial z} \right) + 2\mu_S \left( \frac{\partial \epsilon_{rz}}{\partial r} \right) \tag{18}$$

$$P_\theta = 0 \tag{19}$$

*Surface potential offset.*

The change in the mean-inner potential arising from the bulk flexoelectric effect described in the previous section must be referenced to vacuum to analyze band alignment between two materials in contact. To do so we require the variation in the difference between the vacuum energy and mean-inner-potential (also known as the surface potential offset) with strain.

$$\varphi = \frac{d(\bar{V} - E_{vac})}{d\epsilon} \tag{20}$$

This quantity can be calculated from first-principles [6], measured [7,8], or approximated using analytic atomic/ionic models based upon scattering amplitudes [9,10]. We use the analytic approach to estimate the value of Eq. (20) because the surface potential offset is sensitive to surface details [6,8,10].

A standard analytical approximation introduced by Ibers [9,10] for the mean inner potential with respect to vacuum is

$$\bar{V} = \frac{h^2}{2\pi m e} \frac{1}{\Omega} \sum_i f_{el}^i(0) \tag{21}$$

where $h$ is Plank's constant, $m$ is the electron mass, $e$ is the electron charge, $\Omega$ is the unit cell volume, and $f_{el}^i(0)$ are electron scattering amplitudes. Since we only consider volumetric changes here, it follows that

$$\frac{d\bar{V}}{d\Omega} = -\frac{1}{\Omega^2}\frac{h^2}{2\pi m e} \sum_i f_{el}^i(0) \tag{22}$$

Supplemental Table S1 includes values for the mean-inner potential calculated using Eq. (21) for SrTiO$_3$ and Si and a comparison to experimental measurements from electron microscopy and reflection high energy electron diffraction.

|  | Ionic Factors | Atomic Factors | Measured |
|---|---|---|---|
| SrTiO$_3$ (100) | 22.2 | 15.1 | 13.3 / 14.6 (SrO/TiO$_2$) |
| Si (100) | n/a | 13.8 | 12.1 |

Supplemental Table S1. Comparison between MIP calculated using ionic and atomic scattering factors [9,10] and measurements [7,8]. For SrTiO$_3$ measured values are included for SrO and TiO$_2$ bulk (100) terminations.

*Relative deformation potential.*

The last term needed to analyze band alignment at two interfaces during contact is the relative deformation potential of the conduction band and valence band edges. This term describes shifts in the energy of band feature $E_i$ with respect to the mean-inner potential $\bar{V}$ as a function of strain $\epsilon$.

$$D_i = \frac{d(E_i - \bar{V})}{d\epsilon} \tag{23}$$

For this work we calculated this value from first principles. Details on the calculations are included in Section S4.

*Work function difference.*

Together, the above terms (bulk flexoelectric, surface potential offset, relative deformation potential) describe the shift in a particular band feature with respect to the vacuum level in a single material. To incorporate these effects into a band diagram for two materials in contact it is necessary to also account for their work function difference. The work function difference between $SrTiO_3$ and Si is ~0.6 eV [11,12]. Note, the inclusion of work function differences into our model directly connects to contact potential theory [13].

**S2. Hertz theory and its limitations.**

The Hertz theory of elastic contact [14] provides analytical contact strain fields for several simple geometries. There have been numerous attempts to analytically expand the model, but one of the fundamental assumptions of Hertz theory is that both contacting bodies can be treated as elastic half-spaces. From this assumption it follows that normal contact between two chemically identical bodies with different curvatures (e.g. a sphere contacting a flat made of the same material) will have identical strain fields. This can be seen by considering the expressions for Hertzian stress arising from normal contact between two spheres with different radii ($R_1$ and $R_2$) made of the material (same Young's modulus Y and Poisson's ratio $\nu$). In cylindrical coordinates [15,16], the Hertzian stress components are

$$\frac{\sigma_{rr}}{P} = \frac{1}{2}(1-2\nu)\left(\frac{a}{r}\right)^2\left(1-\left(\frac{z}{u^{\frac{1}{2}}}\right)^3\right) + \frac{3}{2}\left(\frac{z}{u^{\frac{1}{2}}}\right)\left(\frac{(1-\nu)u}{a^2+u} + (1+\nu)\left(\frac{u^{\frac{1}{2}}}{a}\right)ArcTan\left(\frac{a}{u^{\frac{1}{2}}}\right) - 2\right) \qquad (24)$$

$$\frac{\sigma_{\theta\theta}}{P} = \frac{1}{2}(1-2\nu)\left(\frac{a}{r}\right)^2\left(1-\left(\frac{z}{u^{\frac{1}{2}}}\right)^3\right) + \frac{3}{2}\left(\frac{z}{u^{\frac{1}{2}}}\right)^3\left(\frac{a^2 u}{u^2+a^2z^2}\right) + \frac{3}{2}\left(\frac{z}{u^{\frac{1}{2}}}\right)\left(\frac{(1-\nu)u}{a^2+u} + (1+\nu)\left(\frac{u^{\frac{1}{2}}}{a}\right)ArcTan\left(\frac{a}{u^{\frac{1}{2}}}\right) - 2\nu\right) \quad (25)$$

$$\frac{\sigma_{zz}}{P} = \frac{3}{2}\left(\frac{z}{u^{\frac{1}{2}}}\right)^3\left(\frac{a^2 u}{u^2+a^2z^2}\right) \quad (26)$$

$$\frac{\sigma_{rz}}{P} = \frac{3}{2}\left(\frac{r z^2}{u^2+a^2z^2}\right)\left(\frac{a^2 u^{\frac{1}{2}}}{a^2+u}\right) \quad (27)$$

where u is given by

$$u = \frac{1}{2}\left((r^2+z^2-a^2) + ((r^2+z^2-a^2)^2 + 4a^2z^2)^{\frac{1}{2}}\right) \quad (28)$$

the contact radius a is given by

$$a^3 = \frac{3}{4}\frac{F R^*}{Y^*} \quad (29)$$

and the mean contact pressure P is

$$P = \frac{3}{2}\frac{F}{a^2} \quad (30)$$

where F is the applied force. These expressions utilize Y* the effective Young's modulus and R* the effective indenter radius which are defined as

$$\frac{1}{Y^*} = \frac{1-\nu_1^2}{Y_1} + \frac{1-\nu_2^2}{Y_2} \quad (31)$$

$$\frac{1}{R^*} = \frac{1}{R_1} + \frac{1}{R_2} \quad (32)$$

Since the stresses given in Eq. (24)-(27) are each a function of applied force (equal for two bodies in equilibrium), effective modulus (equal in two bodies made of the same material), and the effective indenter radius (equal by construction of the model), all the stress components are equal in both bodies. Assuming the material constituting both bodies exhibits linear elasticity, it follows they will have equivalent strain fields, and no net flexoelectric difference.

While it is often an adequate approximation to use Hertz theory, the result that two chemically identical spheres with different radii in normal contact exhibit the same strain fields is not supported by experiment [17]. Instead, normal contact between two chemically identical spheres is found to yield larger strains in the smaller of the two bodies. These experiments also indicate that the strain fields in the larger body largely agree with those predicted by Hertz theory. This asymmetry acts as the driving force for charge transfer between two identical materials in our model, thus it is necessary to go beyond Hertz theory for the purposes of this work. We accomplish this with finite element modeling.

**S3. Finite element modelling parameters, results, and interpolation.**

*Finite element modelling.*

Finite element simulations were performed using Abaqus. CAX4R mesh elements with reduced integration were utilized owing to the axial symmetry of normal, sphere-on-flat contact. Contact was modelled using frictionless, hard-contact with surface-to-surface discretization. YSYMM mechanical boundary conditions were placed on the bottom-most surface (i.e. no Z-displacement or rotation about the R or Z axes were allowed) and displacements were imposed by prescribing a Z-displacement on the top surface. Supplemental Fig. S2 shows a typical mesh used in this work, but tests were performed with different mesh densities and meshing schemes. The potential maps included in the main manuscript utilized $\sim 10^6$ elements in the flat and $\sim 10^5$ elements in the sphere. Materials were assumed to be isotropic and values of 270 GPa (160 GPa) for Young's modulus and 0.24 (0.27) for Poisson's ratio were used for $SrTiO_3$ (Si) [18,19].

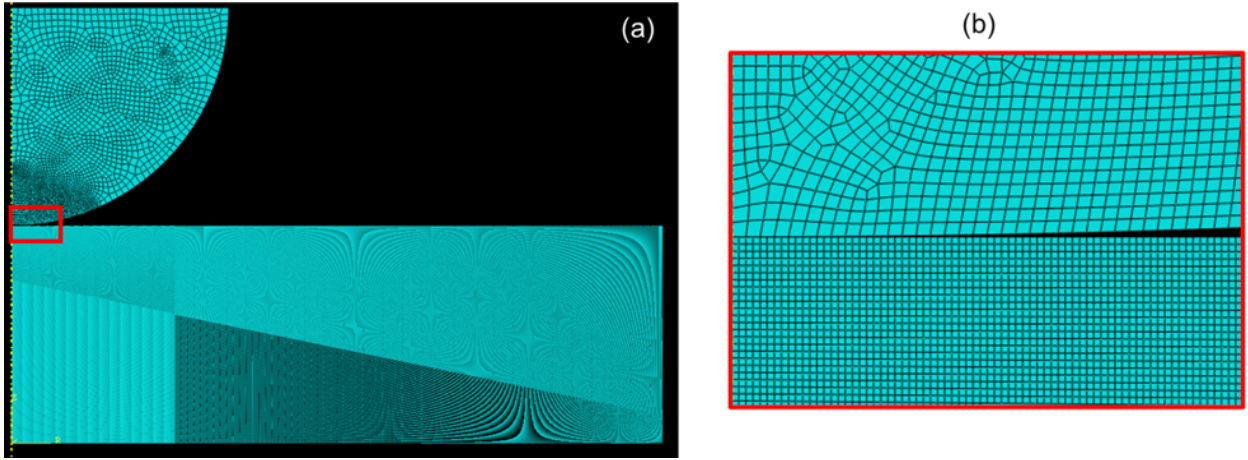

Supplemental Figure S2. (a) An overview of a typical mesh used for Abaqus simulations performed for this work. (b) A magnified view of the vicinity of the contain point showing how the mesh is most dense in this region. ~$10^2$ were present at the contact interface at maximum contact deformation.

*Interpolation.*

Strain fields from Abaqus were then interpolated using a radial basis function approach [20] implemented in Python. Third-order polyharmonic splines were used with a shape parameter of $10^{-5}$. Tests were performed to ensure that the interpolated mesh was adequately dense to minimize errors during differentiation (to get strain gradients) and integration (for the potential). Additionally, multiple basis functions and shape parameter combinations were tested. Spatial derivatives of the interpolated strain fields were used to calculate the non-trivial strain gradient components. Supplemental Figure S3 provides a comparison between an Abaqus strain output and a radial basis function interpolation as a demonstration of the interpolation quality.

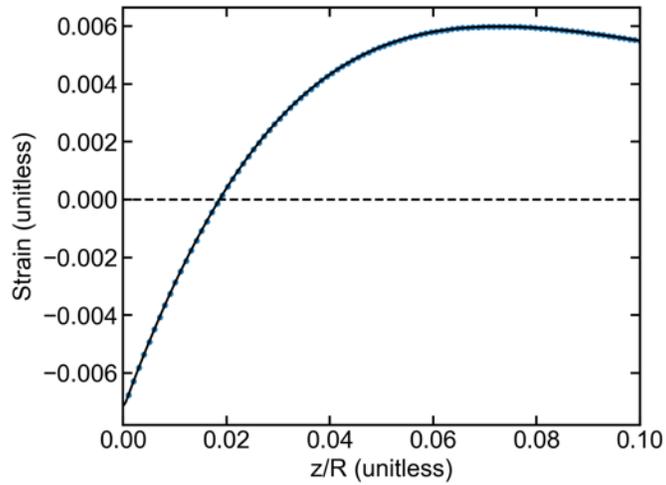

Supplemental Figure S3. Comparison between FEM output (blue points) and RBF interpolation (black line) of $\epsilon_{rr}$ strain at the contact point as a function of depth (z) normalized by the sphere radius (R). The strains are in a $SrTiO_3$ flat in contact with a $SrTiO_3$ sphere with a contact pressure of 8 GPa.

*Deviations from Hertz Theory.*

As shown in the example given in Supplemental Fig. S4, the FEM simulations properly capture that two bodies made of the same material, but with different geometries, have different strain fields in contact. We find that the smaller of the two bodies has larger strain within the vicinity of the contact area (z < 0.1 R, where R is the indenter radius) and the larger body has strain fields that generally match those predicted by Hertz theory. In both cases, strain fields predicted by Hertz theory are recovered sufficiently far from the contact point.

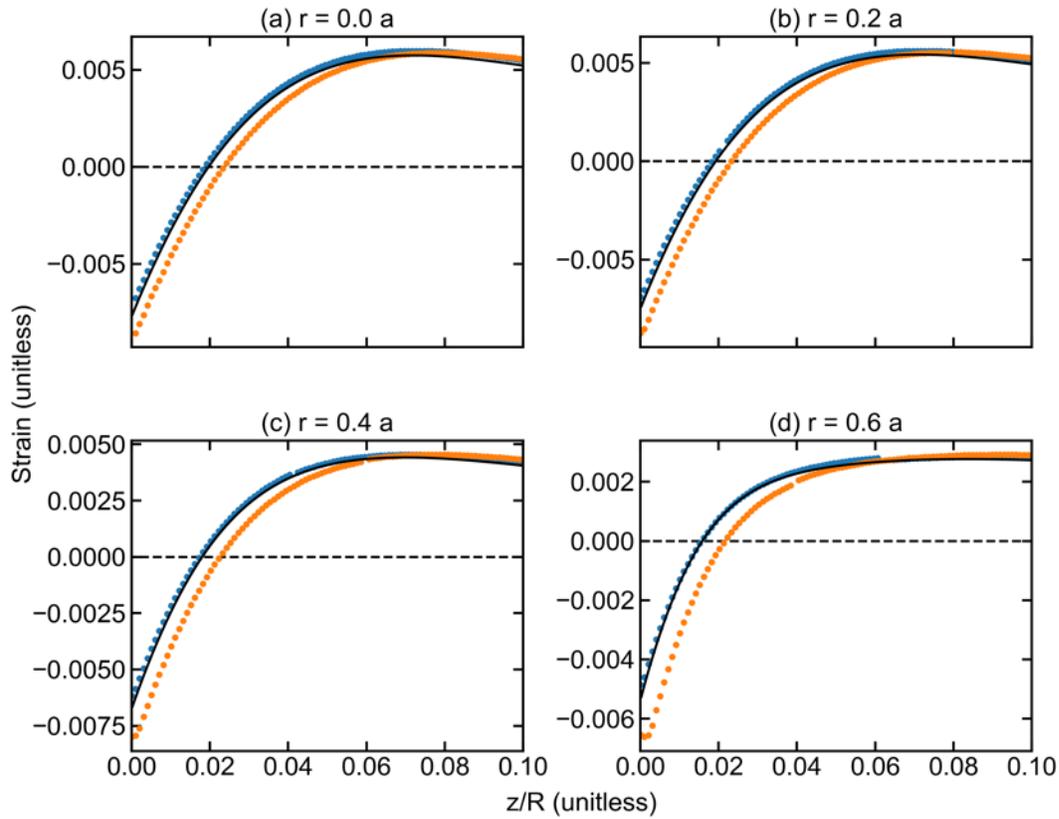

Supplemental Figure S4. $\epsilon_{rr}$ strain for sphere-on-flat contact between two SrTiO$_3$ bodies with a contact pressure of 8 GPa. Orange and blue correspond to strains in the sphere and flat, respectively. The black line indicates the Hertz theory solution. Radial distances (r) are given in terms of the contact radius (a) and axial distance from the point of contact (z) is normalized by the sphere radius (R).

*Strain gradients, flexoelectric polarization, and $\Delta \bar{V}^{FxE}$.*

Supplemental Figures S5 and S6 provide examples of the components of strain gradient in a flat body arising from sphere-on-flat contact. The strain gradients that couple to the radial component of polarization are shown in Supplemental Figure S5 and those that couple to the axial component of polarization are shown in Supplemental Figure S6. The polarization components themselves are shown in Supplemental Figure S7.

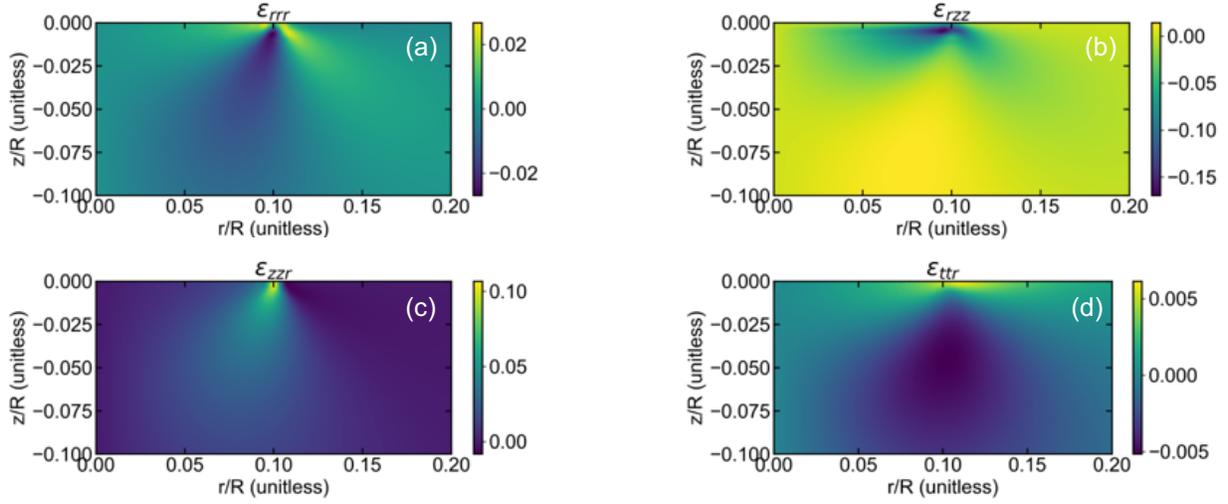

Supplemental Figure S5. Strain gradients in a flat indented by a sphere which couple to radial polarization. Values are given in units of 1/mm. Strain gradients are symmetrized and use the notation $\epsilon_{ijk} = \frac{\partial \epsilon_{ij}}{\partial x_k}$. Simulations correspond to the material properties of SrTiO$_3$ with a contact pressure of 8GPa.

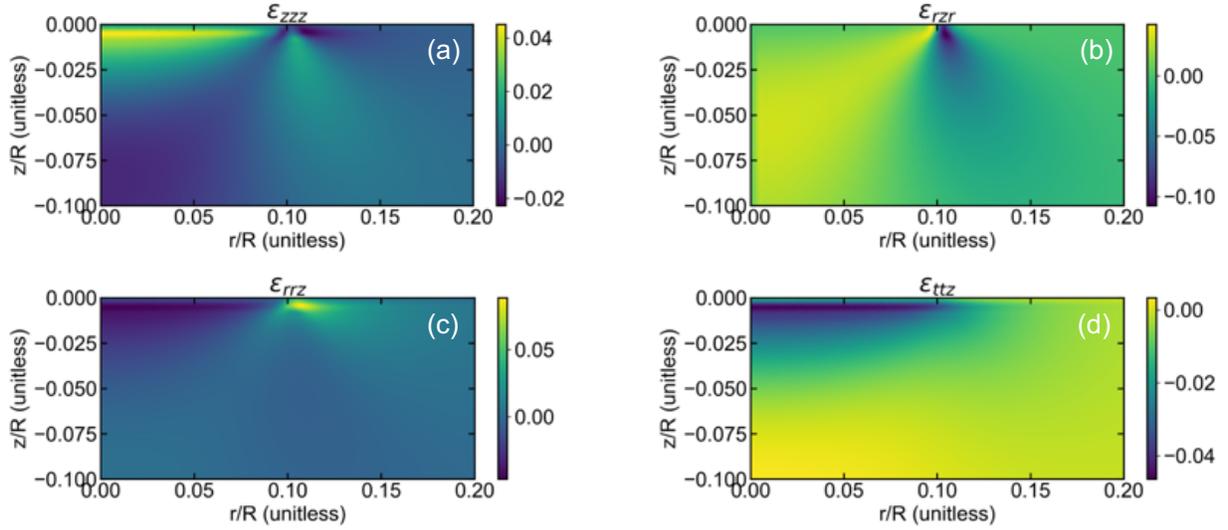

Supplemental Figure S6. Strain gradients in a flat indented by a sphere which couple to axial polarization. Values are given in units of 1/mm. Strain gradients are symmetrized and use the notation $\epsilon_{ijk} = \frac{\partial \epsilon_{ij}}{\partial x_k}$. Simulations correspond to the material properties of SrTiO$_3$ with a contact pressure of 8GPa.

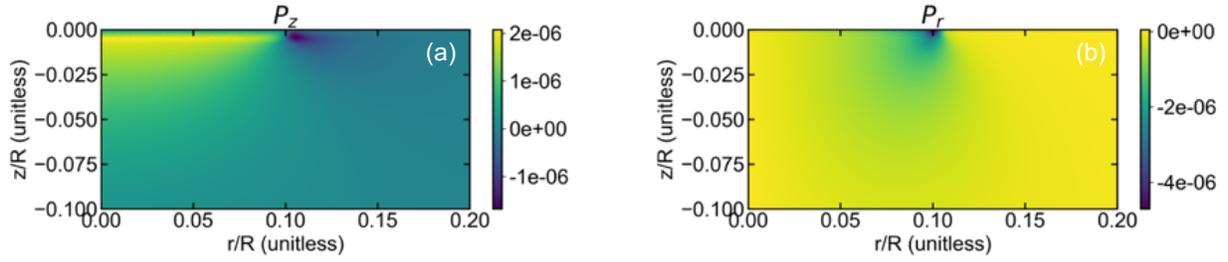

Supplemental Figure S7. (a) Axial and (b) radial polarization in a flat indented by a sphere. Values are given in units of $C/m^2$. Simulations correspond to the material properties of $SrTiO_3$ with a contact pressure of 8GPa.

As described in Section S1, flexoelectric polarization components are used to calculate the change in the mean-inner potential according to Eq. (12). The results of this process are exemplified in Supplemental Figure S8 for sphere-on-flat contact where both bodies are made of $SrTiO_3$.

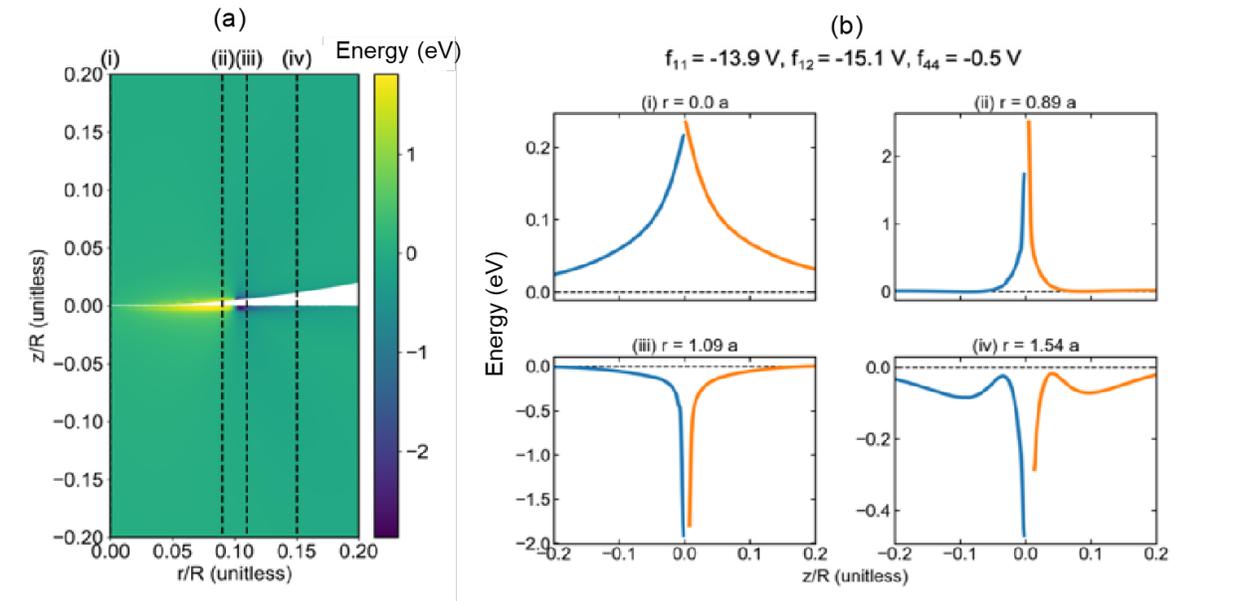

Supplemental Figure S8. Change in the mean-inner-potential for sphere-on-flat contact. Panels (i) - (iv) show the change in the mean-inner potential as a function of depth (normalized by the indenter radius R) at different radial distances (in units of the contact radius a) from the contact point as defined in (a). Simulations correspond to the material properties of $SrTiO_3$ with a contact pressure of 8GPa.

*Volumetric strain effects.*

It is necessary to calculate the volumetric strain $\epsilon_{vol} = \epsilon_{rr} + \epsilon_{\theta\theta} + \epsilon_{zz}$ from the interpolated strains to account for surface potential offset and deformation potential effects described by Eq. (20) and (21). Supplemental Figure S9 shows the volumetric strain corresponding to a SrTiO$_3$ sphere contacting a SrTiO$_3$ flat with a contact pressure of 8 GPa. Note, the differences in Supplemental Figure S9 are not captured with Hertz theory.

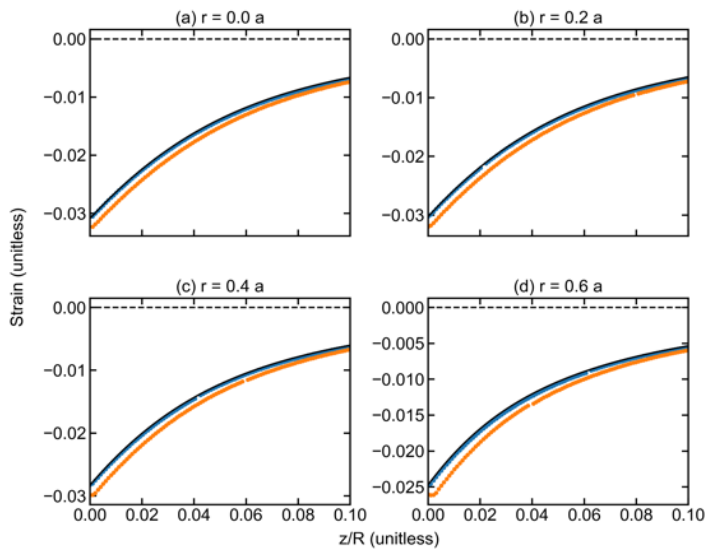

Supplemental Figure S9. Volumetric strain from normal contact between a SrTiO$_3$ sphere (orange) and SrTiO$_3$ flat (blue) compared to Hertz theory strain fields (black). (a) – (d) show the volumetric strain as a function of depth (normalized by the indenter radius R) at different radial distances (in units of the contact radius a) from the contact point. Simulations correspond to a contact pressure of 8GPa.

**S4. Density functional theory calculations.**

*Lattice parameter optimization.*

As a cross-check for parameters in the analysis some density functional theory (DFT) calculations were performed. All DFT calculations were performed with the all-electron augmented plane wave + local orbitals WIEN2K code [21] using the LDA functional for the exchange-correlation term [22]. Muffin-tin radii of 2.44, 1.81, 1.64, and 2.00 au were used for the Sr, Ti, O, and Si atoms, respectively. Optimized lattice parameters were calculated using a plane wave extension value of 6.0 (7.0), k-mesh of 8x8x8 (10x10x10), an energy cut-off of -6 Ry (-8 Ry), Pm-3m (Fd-3m) symmetry for SrTiO3 (Si), a Mermin functional at room temperature, and convergence criteria of $10^{-5}$ e and $10^{-4}$ Ry. This yielded lattice parameters of 7.29 bohr and 10.21 bohr for SrTiO3 and Si.

*Flexoelectric coefficients.*

For this paper we utilize bulk flexoelectric coefficients calculated by Hong and Vanderbilt [23]. We have reproduced these values in Supplemental Table S2 for convenience. Test calculations in WIEN2K yielded comparable results.

|  | $\mu_{1111}$ (nC/m) | $\mu_{1221}$ (nC/m) | $\mu_{1122}$ (nC/m) |
| --- | --- | --- | --- |
| SrTiO3 | - 36.9 | - 1.4 | - 40.2 |
| Si | - 1.3 | - 0.4 | - 0.4 |

Supplemental Table S2. Density functional calculations of flexoelectric coefficients from Ref. [23]. $\mu$ correspond to symmetrized flexoelectric tensor coefficients in units of nC/m. All values correspond to the $E = 0$ boundary condition.

*Hydrostatic deformation potentials.*

Hydrostatic deformation potentials were calculated using finite differences from DFT calculations on bulk SrTiO3 and Si structures with uniform volume expansions/compressions of +/-0.1% about the optimized lattice constant. From these calculations, the hydrostatic deformation potential $D_i$, where i is some band feature such as the conduction band minimum, were calculated according to

$$D_i \approx \frac{(E_i - \bar{V})^+ - (E_i - \bar{V})^-}{3\,(\epsilon^+ - \epsilon^-)} \qquad (34)$$

$E_i$ refers to the energy of the conduction minimum (or valence band maximum) and $\bar{V}$ refers to the MIP in the expanded (+) and compressed (-) strained structures. The results of these calculations are included in Supplemental Table S3. Also included in Supplemental Table S3 are the absolute deformation potentials for these two band structure features in the presence of a [100] longitudinal acoustic phonon. The absolute deformation potential is given by the sum of the hydrostatic deformation potential and the relevant flexocoupling voltages, in this case $f_{1111} = \frac{\mu^E_{1111}}{\chi\, \varepsilon_0}$. These values are included to demonstrate the good agreement between our calculations and the literature [4,5].

|  | $D_{VB}$ (eV) | $D_{CB}$ (eV) | $A_{VB}$ (eV) | $A_{CB}$ (eV) |
|---|---|---|---|---|
| SrTiO$_3$ | - 15.6 | - 17.2 | 3.8 | 2.2 |
| Si | - 11.9 | - 10.2 | 1.3 | 3.0 |

Supplemental Table S3. Hydrostatic ($D_i$) and absolute ($A_i$) deformation potentials for the conduction band minimum and valence band maximum in SrTiO$_3$ and Si. The absolute deformation included here corresponds to strains associated with a [100] longitudinal acoustic phonon. Values are given in eV.

## S5. Band bending as a function of pressure.

It has been shown [24] that

$$V \sim f \left(\frac{F}{R^2 Y}\right)^{\frac{1}{3}} \quad (35)$$

where V is the electrostatic potential at the point of contact between a sphere and flat, f is the effective flexocoupling voltage, F is the applied force, R is the effective indenter radius, and Y is the effective modulus. In its current form, Eq. 35 is useful for a specific experiment (e.g. applying a force using an indenter with a known radius), but for the present purposes it is useful to generalize this expression. From the relationship between force (F), contact radius (a), indentation depth (d), effective indenter radius (R), and effective modulus (Y) given by Hertz theory [14,16]:

$$a^2 = R\,d \tag{36}$$

and

$$F = \frac{4}{3} Y\, R^{\frac{1}{2}} d^{\frac{3}{2}} \tag{37}$$

It follows that

$$F \sim Y\, \frac{a^3}{R} \tag{38}$$

Therefore,

$$V \sim f\left(\frac{F}{R^2 Y}\right)^{\frac{1}{3}} \sim f\left(\frac{a}{R}\right) \tag{39}$$

This can also be written in terms of contact pressure. Using the above equations in addition to the definition of maximum contact pressure,

$$P = \frac{3}{2\pi} \frac{F}{a^2} \sim Y \frac{a}{R} \tag{40}$$

from which it follows that

$$V \sim f\left(\frac{a}{R}\right) \sim f\left(\frac{P}{Y}\right) \tag{41}$$

This analysis maps the effects of different combinations of forces and indenter radii for a particular material onto a single curve. Supplemental Figure S10 compares changes in the mean-inner potential from bulk flexoelectricity at the surface of a SrTiO$_3$ flat contacted by a sphere as functions of a/R and P. This demonstrates large, inhomogeneous band bending (~ ±1 V) can be expected near the contact area. Depending on the flexoelectric properties, the bands can bend in opposite directions inside and outside the contact radius. While nanoscale features ~10 nm on stiff materials with elastic moduli ~100 GPa require forces ~10 nN to exhibit the band bending effects shown in Supplemental Fig. S10, they are also possible in macroscopic features provided proportionately larger forces are applied. In general, an order-of-magnitude upper limit for the maximum contact pressure should be the materials' hardness [25] (e.g. ~8 GPa for $SrTiO_3$ [26]).

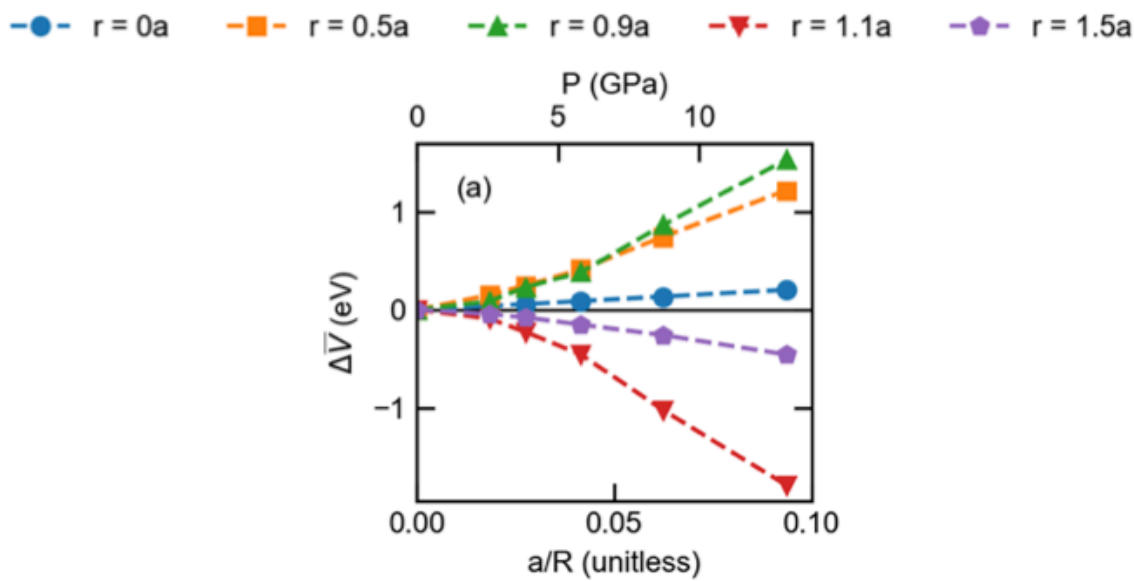

Supplemental Figure S10. Changes in the mean inner potential from flexoelectricity at the surface of a $SrTiO_3$ flat contacted by a sphere as a function of the ratio of the contact radius to indenter radius (a/R) and contact pressure (P). Colors/symbols correspond to different radial distances from the contact point.

**S6. Variations with flexoelectric coefficient.**

$SrTiO_3$ and Si possess large differences in the relative value of their bulk flexoelectric coefficients leading to qualitative differences in contact band bending profiles (Supplemental Table S2). To isolate the effects of the relative size of the flexoelectric coefficients, we calculate the change in the mean inner potential owing to the bulk flexoelectric effect in three fictitious materials which have two out of three flexoelectric coefficients set to zero. This analysis is carried out in a flat contacted by a sphere with a contact pressure of 8 GPa. Supplemental Figure S11 shows differences in the change in the mean-inner potential for (a) $\mu_L = -10\frac{nC}{m}, \mu_T = \mu_S = 0$, (b) $\mu_T =$ $-10\frac{nC}{m}, \mu_L = \mu_S = 0$, and (c) $\mu_S = -10\frac{nC}{m}, \mu_L = \mu_T = 0$. This analysis suggests that shear and transverse contributions to the MIP dominate if the flexoelectric coefficients are of comparable magnitude.

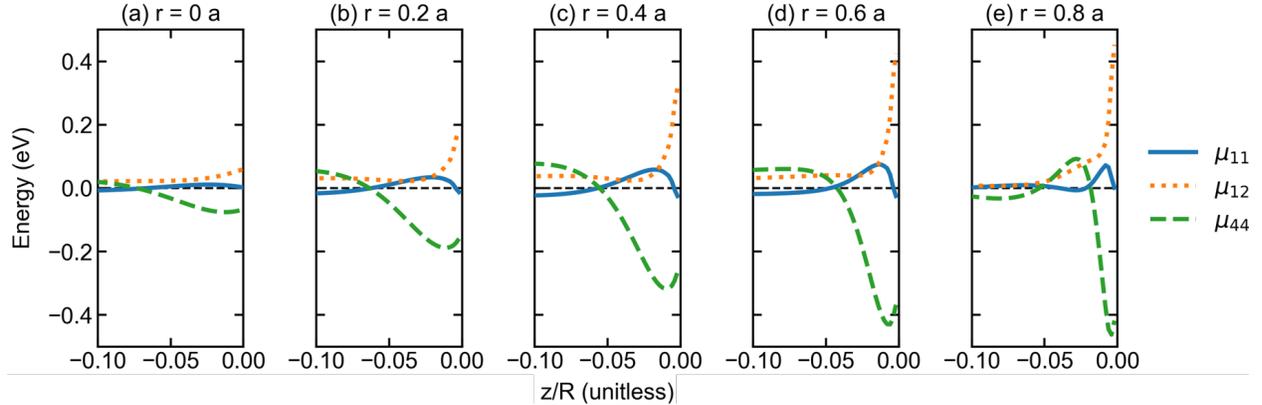

Supplemental Figure S11. Change in the mean-inner potential of a flat contacted by a sphere owing to the bulk flexoelectric effect for three fictitious materials which have $\mu_L = -10\frac{nC}{m}, \mu_T = \mu_S = 0$ (blue, solid), $\mu_T = -10\frac{nC}{m}, \mu_L = \mu_S = 0$ (orange, dotted), and $\mu_S = -10\frac{nC}{m}, \mu_L = \mu_T = 0$ (green, dashed). (a)-(e) shows the change in the mean-inner potential as a function of depth (normalized by the indenter radius R) at different radial distances (in units of the contact radius a) from the contact point. Calculations assume a contact pressure of 6 GPa and elastic properties of SrTiO$_3$.

**S7. Band bending outside the contact radius.**

The main manuscript of this Letter focused on tribocharging mechanisms involving the exchange of electrons or holes between two contacting materials. Therefore, the relevant contact-induced band bending was within the contact radius. For charge transfer mechanisms involving ion exchange, the relevant band bending is that which occurs outside the contact radius. Here we show band bending profiles for radii greater than the contact radius for contact between a Si sphere and SrTiO$_3$ flat (Supplemental Fig. S12) and a SrTiO$_3$ sphere and SrTiO$_3$ flat (Supplemental Fig. S13).

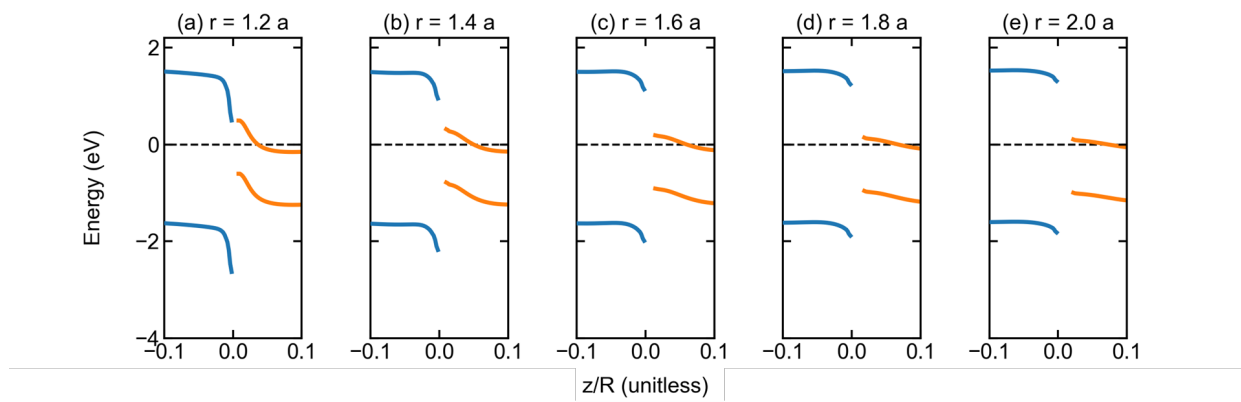

Supplemental Figure S12. Sphere-on-flat contact band diagrams for a Si sphere (orange) on a SrTiO$_3$ flat (blue) including flexoelectric, deformation potential, and surface potential offset effects, and a work function difference of 0.6 eV. (a)-(e) show the conduction and valence band edges as a function of depth (normalized by the indenter radius R) at different radial distances (in units of the contact radius a) from the contact point. The unstrained Fermi level of each material is assumed to be at its band gap center and zero energy is taken to be the unstrained SrTiO$_3$ Fermi level. Calculations assume material parameters given in the main text with a contact pressure of 6 GPa.

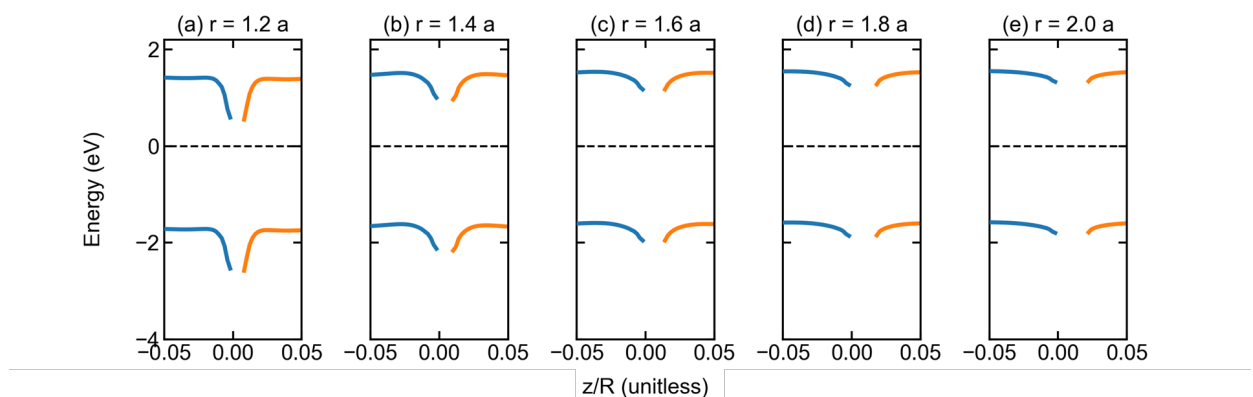

Supplemental Figure S13. Sphere-on-flat contact band diagrams for a SrTiO$_3$ sphere (orange) on SrTiO$_3$ flat (blue) including flexoelectric, deformation potential, and surface potential offset effects. (a)-(e) show the conduction and valence band edges as a function of depth (normalized by the indenter radius R) at different radial distances (in units of the contact radius a). The unstrained Fermi level is assumed to be at the band gap center and taken to be zero energy. Calculations assume material parameters given in the main text with a contact pressure of 8 GPa.

## S8. Connections with other experimental results.

Table S4 includes additional connections between the model presented in this Letter and experiments that were not discussed in the main manuscript. We exclude points already made in Ref. [24].

| Phenomenon | Description | Connection to Flexoelectric Model | Ref. |
|---|---|---|---|
| Velocity Dependence | Investigations of triboelectricity in particles typically measure the velocity-dependence of impact charge (the amount of charge transferred from impact for an initially uncharged sample) and equilibrium charge (the amount of initial charge on a particle needed so that no charge transfer from impact occurs). The velocity dependence of impact charge is attributed to changes in contact area. Equilibrium charge is found to be velocity-independent and proportional to the contact potential difference. | The velocity dependence of impact charge is included in our model because it describes how band bending, the charge transfer driver, increases with contact area. | [27] |
| Dielectric Constant | The literature on the role of the dielectric constant is conflicting: some experiments indicate tribocharge density is linearly proportional to the dielectric constant and others suggest the dielectric constant has no influence. | There is a dielectric constant ($\epsilon$) dependence in the flexoelectric potential term of our model through the flexocoupling coefficient $f = \mu/\epsilon$. While $\epsilon$ is a bulk property, the detailed band bending will depend on the flexoelectric coefficient which is surface-sensitive. This surface-sensitivity could be the source of the discrepancy in the literature. | [28-30] |
| Breakdown Voltage | Dielectric breakdown strength is reduced by contact pressure. | Our model shows that this occurs because of pressure-dependent band bending. | [31] |
| Work function dependence | The tribocharge density is proportional to the work | The work function difference between two materials is built into our model, however we | [29,30] |

| | function difference between two contacting materials. | show that band bending from deformations is also an important consideration. | |
| --- | --- | --- | --- |
| Fracto-luminescence | Light emission is known to occur from the fracture of materials. | Our model describes how elastic deformation increases trap state occupation. The large flexoelectric fields associated with fracture can induce light emission associated with recombination of charge trapped in in-gap states. | [32-34] |

Supplemental Table S4. Additional connections between our model and experiments.